\documentclass{ar-1col}
\usepackage[numbers]{natbib}
\usepackage{url}

\jname{An article prepared for Annu. Rev. Condens. Matter Phys.}
\jvol{8}
\jyear{2017}

\begin{document}

\markboth{B. Yan and C. Felser}{Topological Materials}

\title{Topological Materials: Weyl Semimetals}
\author{ Binghai Yan and Claudia Felser
\affil{Max Planck Institute for Chemical Physics of Solids, Dresden, 01187 Germany; emails: yan@cpfs.mpg.de, felser@cpfs.mpg.de }}
\begin{abstract}
Topological insulators and topological semimetals are both new classes of quantum materials, which are characterized by surface states induced by the topology of the bulk band structure. Topological Dirac or Weyl semimetals show linear dispersion around nodes, termed the Dirac or Weyl points, as the three-dimensional analogue of graphene.
We review the basic concepts and compare these topological states of matter from the materials perspective with a special focus on Weyl semimetals. The TaAs family is the ideal materials class to introduce the signatures of Weyl points in a pedagogical way, from Fermi arcs to the chiral magneto-transport properties, followed by the hunting for the type-II Weyl semimetals in WTe$_2$, MoTe$_2$ and related compounds. Many materials are member of big families and topological properties can be tuned. As one example we  introduce the multifuntional topological materials, Heusler compounds, in which both topological insulators and magnetic WSMs can be found. Instead of a comprehensive review, this article is expected to serve a helpful introduction and summary by taking a snapshot of the quickly expanding field. 
\end{abstract}

\maketitle


\tableofcontents

\url{www.annualreviews.org/doi/abs/10.1146/annurev-conmatphys-031016-025458}

\section{INTRODUCTION}

The insulators are know to be nonconducting because of a finite energy gap that separates the conduction and valence bands. Differences between different insulators were considered only quantitatively such as the band dispersion and the energy gap size. 
Over the last decade, it has been recognized that insulators can be further classified into different classes according to the topology of their band structures~\cite{Qi2011RMP,Hasan:2010ku,Maciejko2011,Hasan2011}. 
For instance, the usual ordering conduction and valance bands of an ordinary insulator can be inverted by strong spin-orbital coupling (SOC), leading to a topological insulator (TI) (see \textbf{Figure 1a}). The inverted bulk band structure topologically gives rise to metallic surface states. Therefore, a TI is characterized by gapless surface states inside the bulk energy gap. These surface states commonly exhibit a Dirac cone type dispersion, in which spin and momentum are locked-up and perpendicular to each other. 
TIs have been observed in many materials~\cite{Yan2012rpp,Ando2013} such as HgTe~\cite{bernevig2006d,koenig2007} and Bi$_2$Se$_3$~\cite{Zhang2009,xia2009}. 
Interestingly, similar to the surface states of TIs, topological surface states  also exist on the surface of many exotic semimetals called Weyl semimetals (WSMs)~\cite{Wan2011,volovik2003universe,Balents2011,Burkov2011,Hosur:2013eb,Vafek2014,WitczakKrempa2014}. Identified by topological Fermi arcs on the surface and chiral magnetic effects in the bulk, WSMs have expanded the repertoire of exotic topological states.

\begin{figure}[h]
\includegraphics[width=5 in]{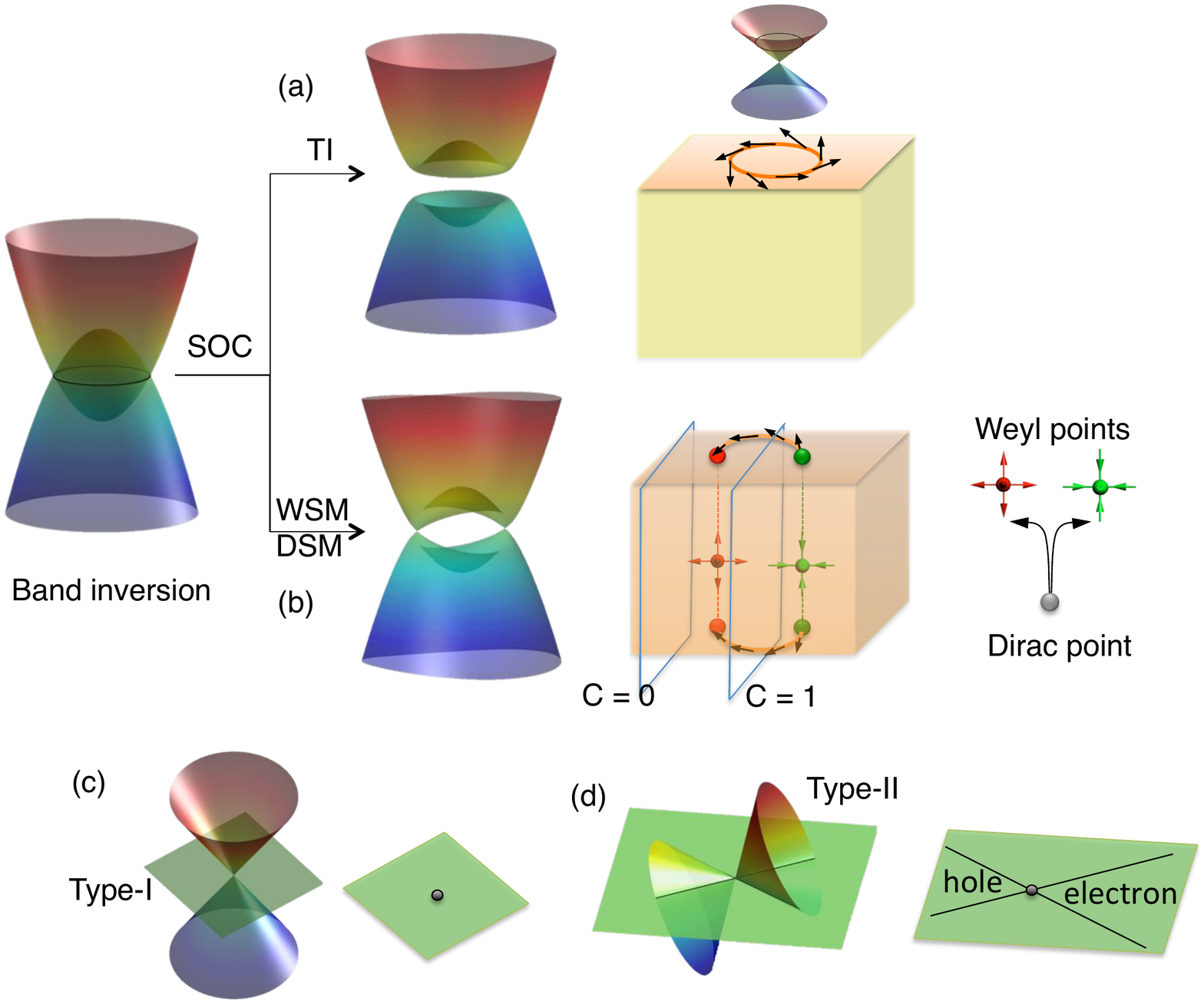}
\label{Figure1}
\caption{The topological insulator (TI) and Weyl semimetal (WSM) or Dirac semimetal (DSM). The topology of both a TI and a WSM/DSM originates from similar inverted band structure. (a) The spin-orbit coupling (SOC) opens a full gap after the band inversion in a TI, giving rise to metallic surface states on the surface. (b) In a WSM/DSM, the bulk bands are gapped by SOC in the 3D momentum space except at some isolating linearly crossing points, namely Weyl points/Dirac points, as a 3D analogue of graphene. Due to the topology of the bulk bands, topological surface states appear on the surface and form exotic Fermi arcs. In a DSM all bands are doubly degenerate while in a WSM the degeneracy is lifted due to  the breaking of the inversion symmetry or time-reversal symmetry, or both. (c) The type-I WSM. The Fermi surface (FS) shrinks to zero at the Weyl points when the Fermi energy is sufficiently close to the Weyl points. (d) The type-II WSM. Due to the strong tilting of the Weyl cone, the Weyl point acts as the touching point between electron and hole pockets in the FS.}
\end{figure}

In 1929, Hermann Weyl demonstrated the existence of massless fermion in the Dirac equation~\cite{Weyl1929}, which was later called the Weyl fermion. In the standard model, all fermions are Dirac fermions, except possibly neutrinos that present the chirality. However, neutrinos were later found to be massive and excluded from Weyl fermions. Weyl fermions have remained undiscovered until very recently in condensed-matter systems~\cite{Weng2015,Huang2015,Xu2015TaAs,Lv2015TaAs,Yang2015TaAs}. 
In solid-state band structures, Weyl fermions exist as low-energy excitations of the Weyl semimetal, in which bands disperse linearly in three-dimensional (3D) momentum space through a node termed a Weyl point. The band structure of a WSM originates from the band inversion in proximity to a TI (see \textbf{Figure 1})~\cite{murakami2007}. 
The Berry curvature is a quantity that can characterize the topological entanglement between conduction and valence bands, which is equivalent to a magnetic field in the momentum space. The Berry curvature becomes singular at Weyl points that act as monopoles~\cite{volovik2003universe} in the momentum space with a fixed chirality: such a Weyl point can be a source (``+'' chirality) or a sink (``--'' chirality) of the Berry curvature. These Weyl points always appear in pairs~\cite{Nielsen1981a,Nielsen1981b}; otherwise, the Berry flux becomes divergent. The WSM requires the breaking of either the time-reversal symmetry (TRS) or the lattice inversion symmetry. When the TRS and inversion symmetry coexist, a pair of degenerate Weyl points may exist, leading to the related Dirac semimetal (DSM) phase~\cite{murakami2007,Young2012,Wang2012,Wang2013}. In other words, a DSM can be regarded as two copies of WSMs. At the critical point during the transition from a TI to a normal insulator, the conduction and valence band touching points are the 3D Dirac points or Weyl points~\cite{murakami2007}, which depends on whether the inversion symmetry exists or not.

Although such gapless band touching has long been known, its corresponding topological nature has been appreciated only recently~\cite{Wan2011}.
Imagine a single pair of Weyl points in a WSM where the TRS is naturally broken, as shown in \textbf{Figure 1b}.
The net Berry phase accumulated in the two-dimensional (2D) $k$ plane between a pair of Weyl points induces a nonzero Chern number $C=1$ with a quantized anomalous Hall effect (AHE) while the Berry phase is zero in other planes with $C=0$.
Therefore, the  anomalous Hall conductivity of the 3D material correspond to the quantized value ($e^2/h$) scaled by the separation of these two Weyl points~\cite{Xu2011,Yang2011QHE,Burkov:2011de,Grushin2012}. This is a pure topological effect from the band structure, since the bulk Fermi surface vanishes at the Weyl point. 
On the boundary, topological edge states exist at edges of 2D planes with $C=1$ and disappear at edges of other planes, where the separation points between two types of planes are these Weyl points.
Consequently, the Fermi surface (FS), an energy contour crossing the Weyl points, exhibits an unclosed line that starts from one Weyl point and end at the other with the opposite chirality, which is called a Fermi arc.
The Fermi arc is apparently different from the FS of a TI, an ordinary insulator or a normal metal, 
which is commonly a closed loop. Therefore, the Fermi arc offers a strong evidence to identify a WSM by a surface-sensitive 
technique such as angle-resolved photoemission spectroscopy (ARPES).
If TRS exists in a WSM, at least two pairs of Weyl points may exist, where TRS transforms one pair to the other by reversing the chirality.
The Fermi arc still appears, as we will discussed in the following. 
However, the AHE is cancelled between these two pairs of Weyl points that contributes opposite Berry phases. 
Instead, an intrinsic spin Hall effect arises~\cite{Sun2016}, which can be considered as the spin-dependent Berry phase and remains invariant under the time-reversal.

An important consequence of the 3D Weyl band structure is that WSMs display the chiral anomaly effect (see \textbf{Figure 2}). In high-energy physics, the particle number of Weyl fermions for a given chirality is not conserved quantum mechanically in the presence of nonorthogonal magnetic (\textbf{B}) and electric (\textbf{E}) fields (i.e., $\textbf{E}\cdot \textbf{B}$ is nonzero), inducing
a phenomenon known as the Adler-Bell-Jackiw anomaly or chiral anomaly ~\cite{Adler1969,Bell1969}. 
In WSMs, the chiral anomaly is predicted to lead to a negative magnetoresistance (MR) due to the chiral zero modes of Landau levels of the 3D Weyl cones and 
suppressed backscattering of electrons of opposite chirality~\cite{Nielsen1983,Son:2013jz}. 
The negative MR is expect to exhibit the largest amplitude when \textbf{B $\|$ E}, since the  \textbf{B $\bot$ E} component contributes a positive MR due to the Lorentz force.
In addition to the negative MR, the chiral anomaly is also predicted to induce exotic nonlocal transport and optical properties~\cite{Parameswaran2014,Potter2014,Baum2015}.

\begin{figure}[h]
\includegraphics[width=5 in]{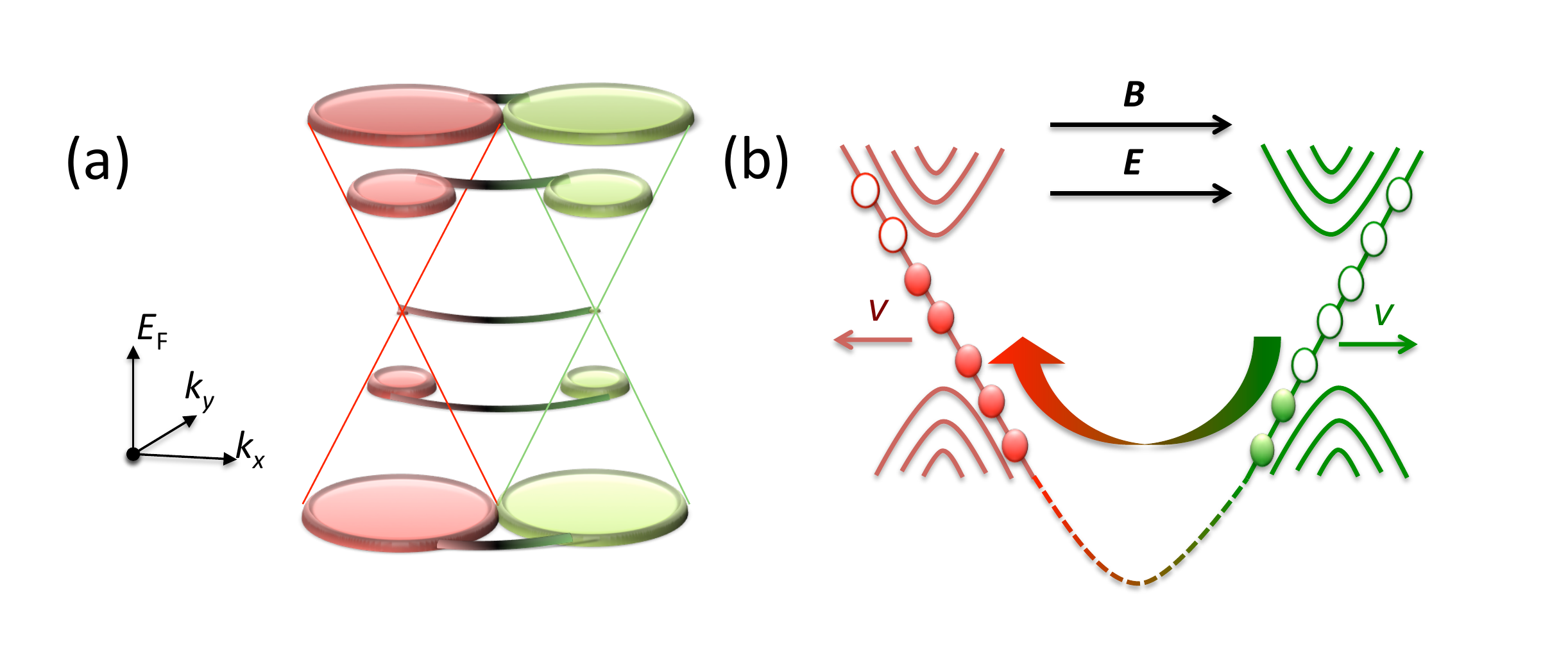}
\label{Figure2}
\caption{Schematics of Fermi arcs and the chiral anomaly effect. (a) Existence of Fermi arcs in the Fermi surface of the surface band structure. A pair of bulk Weyl cones exists as a pair of Fermi pockets at a Fermi energy $E_{\mathrm{F}}\neq0$ or as points at $E_{\mathrm{F}}=0$, where the red (green) color represents the ``+'' (``--'') chirality.  A Fermi arc (thick line) appears on the top or bottom surface to tangentially connect such a pair of Fermi pockets. (b) The chiral anomaly can be simply understood with the zeroth Landau level of a WSM in the quantum limit. The zeroth Landau levels from the``+'' and ``--'' chiral Weyl cones exhibit opposite velocities due to different chirality. Applied electric field leads to imbalance of electron densities in the left-hand and right-hand valleys, breaking the number conservation of electrons at given chirality. This result, however, is not limited to the quantum limit and was recently proposed to induce an anomalous DC current that is quadratic in the field strength in the semi-classical limit.
}
\end{figure}

Many material candidates have been predicted as WSMs, e.g., the pyrochlore iridate Y$_{2}$Ir$_2$O$_7$~\cite{Wan2011}, HgCr$_2$Se$_4$~\cite{Xu2011} and Hg$_{1-x-y}$Cd$_x$Mn$_y$Te~\cite{Bulmash2014}.
However, these candidates have not been experimentally realized thus far. 
In early 2015, four WSM materials -- TaAs, TaP, NbAs, and NbP -- were discovered through calculations~\cite{Weng2015,Huang2015} and the observation of Fermi arcs using ARPES ~\cite{Xu2015TaAs,Lv2015TaAs,Yang2015TaAs}, realizing Weyl fermions for the first time (also in photonic crystals~\cite{Lu2015}). Meanwhile, many efforts have been devoted to their magneto-transport properties~\cite{Huang2015anomaly,Zhang2016ABJ,Shekhar2015,Arnold2016}. These WSM compounds preserve TRS but break crystal inversion. In addition, the DSMs were found to exist in Na$_3$Bi and Cd$_3$As$_2$ by ARPES~\cite{Liu:2014bf,Liu2014Cd3As2,Neupane2014,Borisenko2014}.

WSMs can be classified into type-I, which respects Lorentz symmetry, and type-II, which does not (see {\textbf{Figure 1}})~\cite{Soluyanov2015WTe2}. The TaAs family of WSMs exhibits ideal Weyl cones in the bulk band structure and belongs to the type-I class, i.e., the Fermi surface (FS) shrinks to a point at the Weyl point. More recently, type-II WSMs have been proposed to exist in the layered transition-metal dichalcogenides WTe$_2$~\cite{Soluyanov2015WTe2} and its sister compound MoTe$_2$\cite{Sun2015MoTe2}. Here, the Weyl cone exhibits strong tilting so that the Weyl point appears as the contact point between an electron pocket and a hole pocket in the FS. 
Type-II WSMs are expected to show very different properties from type-I WSMs, such as anisotropic chiral anomaly depending on the current direction and a novel anomalous Hall effect~\cite{Soluyanov2015WTe2}. In addition, the layered nature of these compounds can facilitate the fabrication of devices, making them an ideal platform for the realization of novel WSM applications.

In this article, we will overview recent progress in topological states of matter from the viewpoint of materials. We start from the TaAs family of WSMs by  introducing their surface states and chiral magnetotransport properties. We further follow the recent search for type-II WSMs in MoTe$_2$ and related compounds. Then, we introduce the multifuntional topological materials termed Heusler compounds, in which both topological insulators and TRS-breaking WSMs can be found.

\section{WEYL SEMIMETALS: THE TAAS FAMILY}

\subsection{Bulk and surface states}

\begin{figure}[h]
\includegraphics[width=4 in]{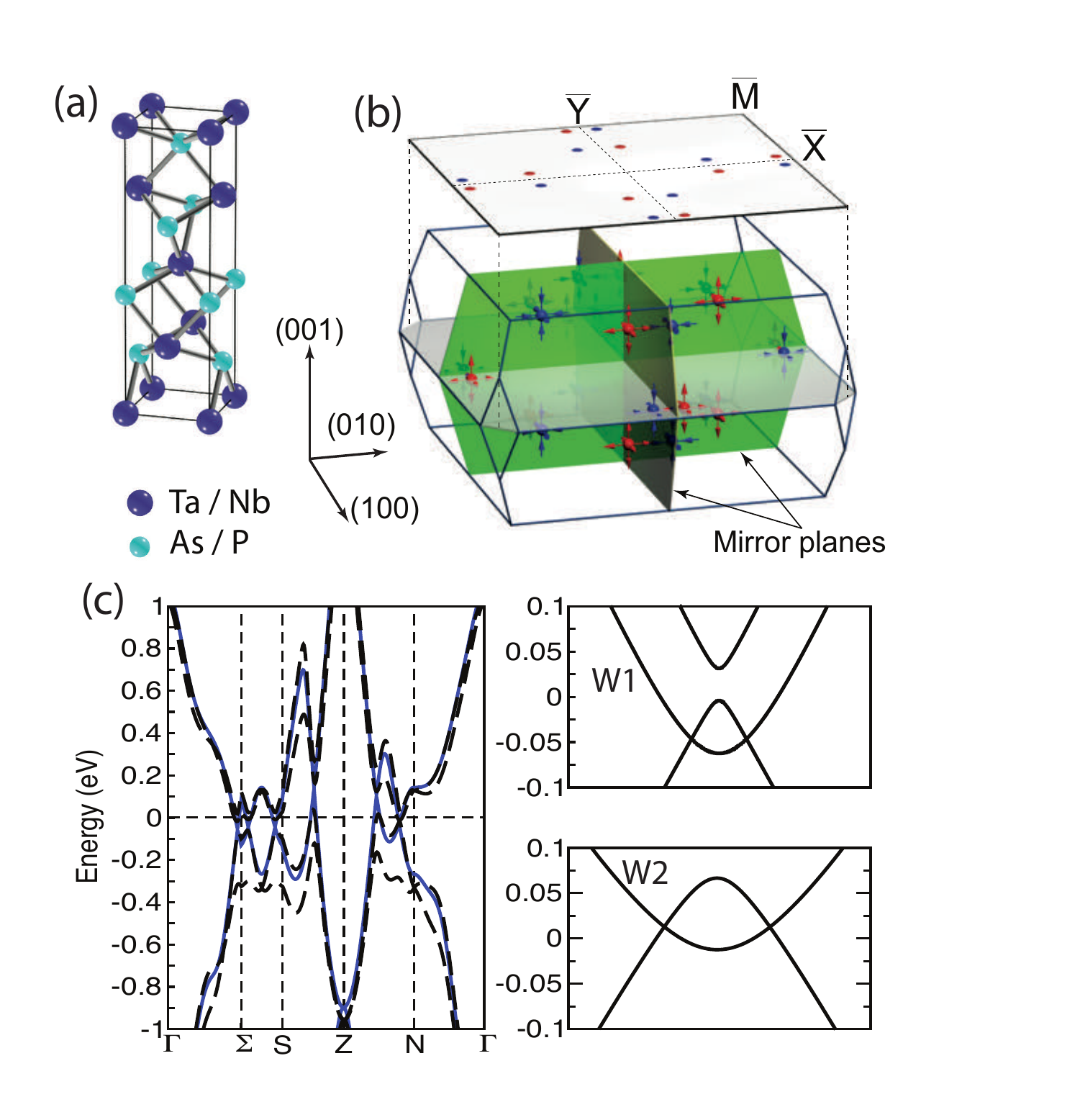}
\label{Figure3}
\caption{Crystal structure and bulk band structure.  (a) The non-centrosymmetric crystal lattice of TaAs-family compounds. (b) The first Brillouin zone with twelve pairs of Weyl points. The red and blue spheres represent the Weyl points with ``+'' and ``-'' chirality, respectively. The arrows surround a Weyl point stand for the monopole-like distribution of the Berry curvature. (c) The bulk band structure along some high symmetry lines. The solid blue lines  and black dashed lines corresponding without and with SOC, respectively. The energy dispersions crossing a pair of W1 and W2 Weyl points are shown in the right panels. The Fermi energy is set to zero.
[Adapted from Liu $et~al$ ~\cite{Liu2016NbPTaP} and Sun $et~al$ ~\cite{Sun2015arc}.]
}
\end{figure}

We first use the TaAs-type compounds as an example to demonstrate the properties of WSMs. Meanwhile, we will answer a simple but essential question in a pedagogical manner: how can we identify a WSM in both theory and experiment? 

(i) Weyl points as the band-crossing points. According to the definition, the conduction and valence bands touch at the Weyl points in the bulk band structure, which is usually obtained from state-of-art $ab~initio$ calculations. Therefore, the first step is to address these gapless points in the 3D Brillouin zone (BZ). Sometimes, it can be challenging because Weyl points usually stay away from high-symmetry lines or planes and their total numbers are not easy to identify. Fortunately, the band structure without including SOC can provide insightful hints for solving this problem. Take TaP as an example (see \textbf{Figure~3}). TaP crystallizes into a noncentronsymmetric tetragonal lattice (space group $I4_1md$, No. 109) that contains two mirror planes, $M_x$ and $M_y$. One can find band crossing at some high-symmetry lines in the band structure when SOC is turned off. These gapless points form nodal rings in the mirror planes in the BZ. When SOC is switched on, the band structure exhibits spin splitting due to the lack of inversion symmetry. Consequently, the nodal lines are all gapped. However, band-touching points exist near the original nodal rings but away from the mirror planes. A single nodal ring evolves into three pairs of gapless points that can be classified into two groups, one pair in the $k_z=0$ plane (labelled W1) and two pairs in the $k_z \approx \pm \pi/c$ planes, where $c$ is the lattice parameter along the $z$ axis. In total, there are twelve pairs of Weyl points, considering four nodal rings inside two mirror planes in the first BZ. Along a line that connects a pair of Weyl points, one can find that the conduction and valence bands indeed cross each other linearly through Weyl points (\textbf{Figure~3c}). We point out that Weyl points W1 and W2 appear at slightly different energies in the band structure. Experimentally, the linear dispersions near two types of Weyl points have been visualized using soft X-ray ARPES, which is sensitive to the bulk states~\cite{Xu2015TaAs,Yang2015TaAs,Lv2015TaAsbulk}.

(ii) Weyl points as monopoles. The topology of Weyl points can be verified through the Berry curvature. The Berry curvature $\bf{\Omega(k)}$ characterizes the wave-function entanglement between the conduction and valence bands and is considered as an effective magnetic field (a pseudo-vector) in the $k$-space, which can be obtained based on the Bloch wave functions from band-structure calculations~\cite[and references therein]{Xiao2010}. The Weyl point behaves as a monopole, i.e., a source or sink of this field. The corresponding magnetic charge, i.e., chirality $\chi$, can be found as an integral over the the FS enclosing the Weyl point, $\chi = \frac{1}{2\pi} \oint_{\mathrm{FS}} \mathbf{\Omega(k)} \cdot d\mathbf{S(k)}$. In this manner, the monopole feature of all Weyl points can be identified for TaP.  Although many Weyl points exist in the BZ, we can easily organize them according to their symmetry. 
Regarding two mirror planes $M_x$ and $M_y$ that reverse the chirality, one Weyl point at the position $(k_x, k_y, k_z)$ with, for instance, chirality $\chi=1$ has three partners: one each at $(-k_x, k_y, k_z)$ and $(k_x, -k_y, k_z)$ with $\chi=-1$ and one at $(-k_x, -k_y, k_z)$ with $\chi=1$. Considering the TRS that preserves the chirality, the above four Weyl points have four time-reversal partners at  $(\pm k_y, \pm k_x, \pm k_z)$. 
Further, these eight Weyl points  at $(\pm k_x, \pm k_y, \pm k_z)$ find another eight partners at $(\pm k_y, \pm k_x, \pm k_z)$ by considering the $C_4$ rotation that also maintains the chirality. Therefore, there are eight W1-type Weyl points since $k_z=0$ and sixteen W2-type Weyl points  since $k_z \neq 0$. 

(iii) Fermi arcs on the surface. A significant consequence of the monopole feature of Weyl points is the existence of Fermi arcs in the FS of the surface states. Consider a simple WSM in which only a pair of Weyl points exists with TRS breaking and slices not containing the Weyl points in the BZ are gapped as 2D insulators. If the slices lie between two Weyl points, the net Berry flux accumulated across the slices leads to Chern insulators with a nonzero Chern number ($C$), which carry chiral edge states. If the slices stay far away from the two Weyl points, they are trivial 2D insulators. Thus, the chiral edge states exist only between these two Weyl points and assemble a topological surface state. When the Fermi energy crosses the Weyl points, a Fermi arc exists to connect the surface projections of two Weyl points (see \textbf{Figure~2}). The Fermi arcs from the top and bottom surfaces  are related through the bulk Weyl points, which can generate fascinating transport and optical phenomena~\cite{Potter2014,Moll2016Cd3As2,Baum2015}. The 3D band structures are shown in \textbf{Figure 2}. When the Fermi energy is away from the Weyl points, the Fermi arc tangentially connects two bulk Weyl pockets. We point out that the Fermi arcs may appear even when two Weyl pockets merge. However, these argument based on the Chern number cannot be applied for TaP because of the existence of TRS, which constrains the Chern number to be zero. Instead, the same topological $Z_2$ index as that of a TI can be defined to identify the Fermi arcs~\cite{Weng2015}. The mirror plane, for example $M_x$ in the BZ that lies between six pairs of Weyl points, is a gapped 2D TI exhibiting a nontrivial $Z_2$ index. Consequently, helical edge states exist, which form Fermi arcs in the surface band structure as well. 

\begin{figure}[h]
\includegraphics[width=4 in]{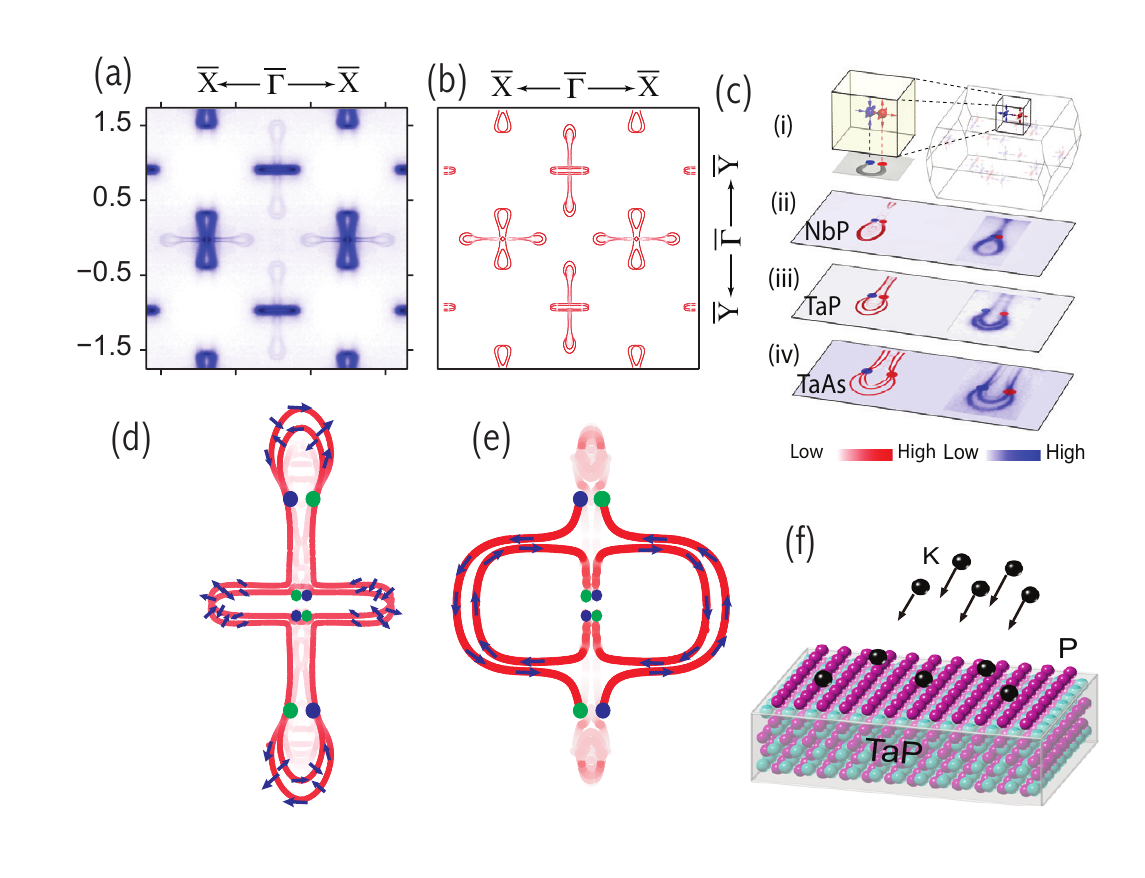}
\label{Figure4}
\caption{Fermi arcs from ARPES and theoretical calculations. The Fermi surface of TaP (a) by ARPES and  (b) by calculations agree very well. 
(c) Evolution of the band structures for NbP, TaP and TaAs, in which SOC increases in order.  (i) Schematic plot (grey curves) connecting them. (ii--iv) Comparison of the calculated (left) Red/blue dots denote the Weyl points of opposite chirality. (d) The Fermi surface around the $\bar{Y}$ point. The spoon-head part corresponds to the Fermi arcs. The bowtie-like middle part is due to trivial surface states. Blue arrows indicate the spin texture. (e) Calculated new Fermi surface after a Liftzshitz transition, which is formed purely by long Fermi arcs. (f) The illustration of the Lifzshitz transition induced by depositing potassium atoms on the TaP surface.
[Adapted from Liu $et~al$ ~\cite{Liu2016NbPTaP} and Sun $et~al$ ~\cite{Sun2015arc}.]
}
\end{figure}

The existence of Fermi arcs is topologically protected by the Weyl points in the underlying bulk. However, the shape and energy dispersion of Fermi arcs is sensitive to the surface. Indeed, it is found that the surfaces terminated by cations (Ta, Nb) and anions (As, P) present very different FSs in $ab~initio$ simulations~\cite{Sun2015arc}. In general, anion-terminated surfaces were usually reported for the as-cleaved surface in ARPES for TaAs
~\cite{Xu2015TaAs,Lv2015TaAs,Yang2015TaAs}, TaP~\cite{Liu2016NbPTaP}, NbAs~\cite{Xu2015NbAs} and NbP~\cite{Liu2016NbPTaP,Belopolski2016NbP}, while anion-terminated surfaces were also observed for TaP~\cite{Xu2016TaP} and NbP~\cite{Souma2015NbP}. The typical FS of the As- or P-terminated surface is shown in \textbf{Figure 4}; it is composed of spoon-like and bowtie-like parts. Two pieces of Fermi arcs appear at the head of the spoon-like region, connecting the projections of W2-type Weyl points. These two arcs exhibit opposite spin texture~\cite{Yang2015TaAs,Sun2015arc,Lv2015TaAsspin,Xu2016TaAsspin}. The separation of W2 Weyl points with opposite chirality and the length of corresponding Fermi arcs are proportional to the strength of SOC, i.e., decreasing in the order of TaAs, TaP, NbAs, and NbP~\cite{Sun2015arc, Liu2016NbPTaP}. 
The bowtie-like part is due to the $p$-orbital dangling bond states of As or P. As demonstrated by calculations, these dangling bonds can be passivated, for instance, by the guest atom deposition, resulting in a Liftzshitz transition of FS that is purely composed of the Fermi arcs~\cite{Sun2015arc}. Additionally, the Fermi arcs can also be visualized in scanning tunneling spectroscopy~\cite{Inoue2016,Batabyal2016}, where the topology of the FS and its spin texture are essential to understand the quasiparticle interference on the surface of WSMs~\cite{Kourtis2016}.

\subsection{Chiral magneto-transport}

The magnetotransport properties have been quickly and widely studied for WSM materials. The linear dispersion and nontrivial Berry phases in the band structure allow the detection of chiral magnetotransport phenomena, large MR  and high mobility as well. Strong quantum oscillations also assists for the reconstruction of the Fermi surface information, to further validate and reveal the physics behind.

\subsubsection{Chiral anomaly and fermiology}
The discovery of WSM materials triggered an experimental search for the exotic quantum phenomenon known as the chiral anomaly in condensed-matter physics. Recently, a negative longitudinal MR, the resistivity change under magnetic field [$\Delta\rho(B)/\rho(0)$] and $\mathbf{B || I}$,
has been reported in two types of WSMs: WSMs induced by time-reversal symmetry breaking, i.e., Dirac semimetals in an applied magnetic field, for example Bi$_{1-x}$Sb$_x$ ($x\approx3 \%$)~\cite{Kim2013}, ZrTe$_5$,~\cite{Li2016ZrTe5}, Na$_3$Bi~\cite{Xiong2015} and Cd$_3$As$_2$~\cite{Li2016Cd3As2},
 and the non-inversion-symmetric WSMs TaAs~\cite{Huang2015anomaly,Zhang2016ABJ}, TaP~\cite{Arnold2016}, NbP~\cite{Wang2015NbP}. All these systems are semimetals with a very high mobility as in classical semimetals such as bismuth and graphite. The chiral anomaly is believed to induce the negative longitudinal MR~\cite{Nielsen1983,Son:2013jz}. However, the negative longitudinal MR does not solely originate from the chiral anomaly effect. For example, it can be induced by the inhomogeneous current distribution in the device~\cite{Arnold2016,Yoshida1976}, a generic 3D metal in the quantum limit~\cite{Goswami2015}, or other effects without necessarily evolving Weyl points~\cite{Chang2015,Ma2015,Zhong2016}. Therefore, a clear verification of the existence of Weyl points in the FS topology is crucial to interpret the MR experiments. 

In the TaAs family of WSMs, two types of Weyl points, W1 and W2, exist at different momentum positions and energies. In the FS, the Weyl electrons generally coexist with topologically trivial normal electrons. In principle, small changes of the Fermi energy ($E_\mathrm{F}$), as induced by doping or defects, can significantly change the Fermi-surface topology owing to the low intrinsic charge-carrier density in semimetals. 
Even in the ideal case of completely stoichiometric and compensated crystals, the Fermi energy does not necessarily cross the Weyl points. 
Therefore, a precise knowledge of $E_\mathrm{F}$ and the resulting Fermi-surface topology is required when linking the negative MR to the chiral-magnetic effect.
Although ARPES studies have shown the existence of Fermi-arc surface states and linear band crossings in the bulk band structure of all four materials, its energy resolution ($> 15~\mathrm{meV}$) is insufficient to confirm the presence or absence of Weyl fermions at the Fermi level. In contrast, quantum-oscillation measurements have the advantage of meV-order resolution of the Fermi level. Quantum oscillations originating from different types of FS pockets were found in magnetization, magnetic torque, and MR measurements performed in magnetic fields at low temperature. These oscillations are periodic with respect to $1/B$. Their frequency ($F$) is proportional to the corresponding extremal FS cross section ($A$) that is perpendicular to \textbf{B} following the Onsager relation $ F = (\Phi_0 / 2\pi^2) A$, where $\Phi_0 = h/2e $ is the magnetic flux quantum and $h$ is the Planck constant.
To reconstruct the shape of the FS, the full angular dependence of the quantum-oscillation frequencies was compared to band-structure calculations~\cite{Arnold2016FS,Arnold2016} . The exact position of $E_\mathrm{F}$ was determined by matching the calculated frequencies and their angular dependence to the experimental ones.

The 3D FSs of TaAs~\cite{Arnold2016FS}, TaP~\cite{Arnold2016}, and NbP~\cite{Klotz2016} have very recently been reconstructed by combining sensitive angle-dependent Shubnikov-de Haas (SdH) and de Haas-van Alphen (dHvA) oscillations with \textit{ab~initio} band-structure calculations, showing an excellent agreement between theory and experiment. In these semimetals, electron and hole pockets coexist near the nodal ring positions in the FS, as shown in (\textbf{Figure 4}). For TaAs, $E_\mathrm{F}$ crosses the upper parts of both W1 and W2 Weyl cones, leading to independent W1 and W2 electron pockets. This type of FS can suppress the scattering between a pair of Weyl valleys with opposite chirality, supporting the existence of the chiral anomaly effect. However, the results for TaP and NbP are not found to be the same. $E_\mathrm{F}$ is located far above the W1 Weyl points and slightly below the W2 Weyl points. Consequently, W1 and W2 Weyl points are swallowed inside large electron and hole pockets, respectively. The disappearance of the individual Weyl pocket, which includes a single Weyl point, quenches the chiral anomaly effect. Instead, the negative MR observed in TaP was further explained as a field-induced current redistribution~\cite{Arnold2016}. With the aim of realizing chiral anomaly in TaP and NbP, electron doping was suggested to shift the Fermi energy in the close vicinity of the W2 Weyl points~\cite{Arnold2016,Klotz2016}.
In addition, the topology of the Fermi surface was found to be robust under external pressure up to $2\sim3$ GPa~\cite{Luo2016,Reis2016}. In NbP the W2 Weyl points were found to move close to the Fermi energy by 3 meV corresponding to a pressure about 3 GPa~\cite{Reis2016}.

\begin{figure}[h]
\includegraphics[width=5 in]{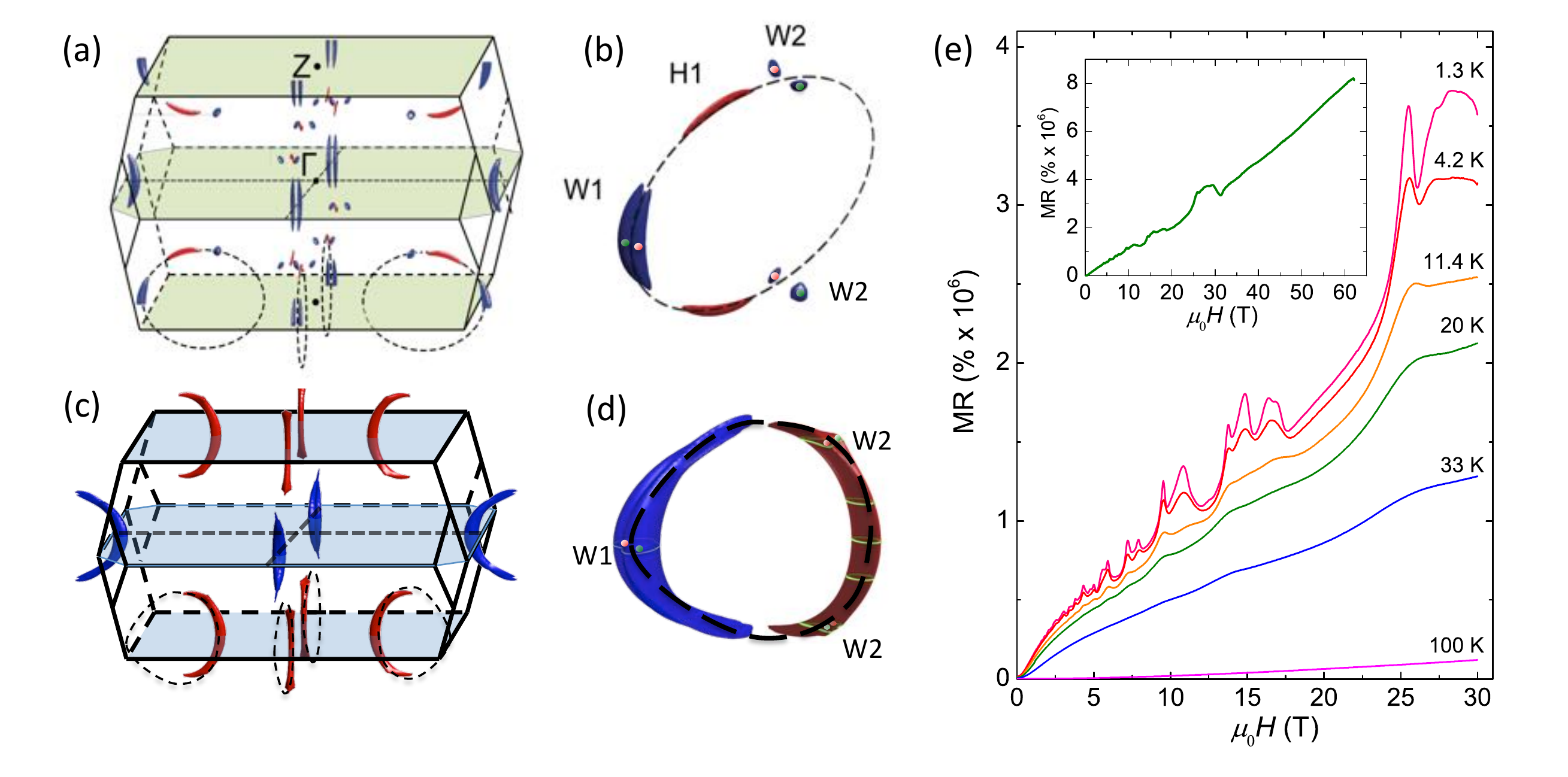}
\label{Figure5}
\caption{Bulk Fermi surfaces constructed by quantum oscillations and calculations. (a), (b) TaAs Fermi surface. Blue and red Fermi surfaces correspond to electron and hole pockets, respectively. The dashed loop represents the nodal ring. The positions of Weyl points are indicated by red/green points.
Each Weyl point (both W1 and W2 types) is independently included inside a single electron pocket, which is active for the chiral anomaly.
(c -- d), TaP Fermi surface. All W1 and W2 Weyl points are included inside a large electron and hole pockets, respectively.
(e) Large transverse magnetoresistance (MR) with Shubnikov-de Hass oscillations measured on NbP.
[Adapted from Shekhar $et~al$ ~\cite{Shekhar2015} and Arnold  $et~al$ ~\cite{Arnold2016,Arnold2016FS}.]
}
\end{figure}

\subsubsection{Large positive magnetoresistance and high mobility} 
WSMs and DSMs usually exhibit very high mobilities that can probably be attributed to the chiral and massless feature of Weyl or Dirac fermions and the high Fermi velocity, as observed in transport experiments with materials such as Cd$_3$As$_2$~\cite{Liang2014,Narayanan2015} and NbP~\cite{Shekhar2015}. 
Generally, semimetals are new platforms to realize a huge transverse MR ($\mathbf{B \bot I})$, an effect that has been pursued intensively in emerging materials in recent years because of its significant potential for application in state-of-the-art information technologies~\cite{Parkin:2003id}.
Electrical transport in a semimetal usually consists of two types of carriers (electrons and holes), leading to large MR when a magnetic field is applied with an electron--hole resonance~\cite{Singleton2001band}. In a simple Hall effect setup, the transverse current carried by a particular type of carrier may be nonzero, although no net transverse current flows when the currents carried by the electrons and holes compensate for each other. These nonzero transverse currents will experience a Lorentz force caused by the magnetic field in the inverse-longitudinal direction. Such a back flow of carriers eventually increases the apparent longitudinal resistance, resulting in a dramatic MR that is much stronger than that in normal metals and semiconductors. Thus, high-purity samples are crucial to realize a balance between electrons and holes and a high carrier mobility ($\mu$), both of which will enhance the positive MR. 

In the TaAs family of WSMs, NbP~\cite{Shekhar2015} was first reported to exhibit an extremely large positive MR of 850,000\% at 1.85~K (250\% at room temperature) in a magnetic field of up to 9 T, which increases linearly up to 60 T without any signs of saturation (see \textbf{Figure 5e}). Ultrahigh carrier mobility accompanied by strong SdH oscillations was also observed. A perfect electron-hole compensation was identified on the FS~\cite{Klotz2016}, explaining the large MR observed. Similar behaviors were also observed in other compounds of this family~\cite{Ghimire2015NbAs,Huang2015anomaly,Zhang2016ABJ,Wang2015NbP,Luo2015,Moll2015}. Therefore, this family of WSMs presents profound examples of materials that combine topological and conventional electronic phases with intriguing physical properties resulting from their interplay. 

\section{WEYL SEMIMETALS: The TYPE-II}

\begin{figure}[h]
\includegraphics[width=2 in]{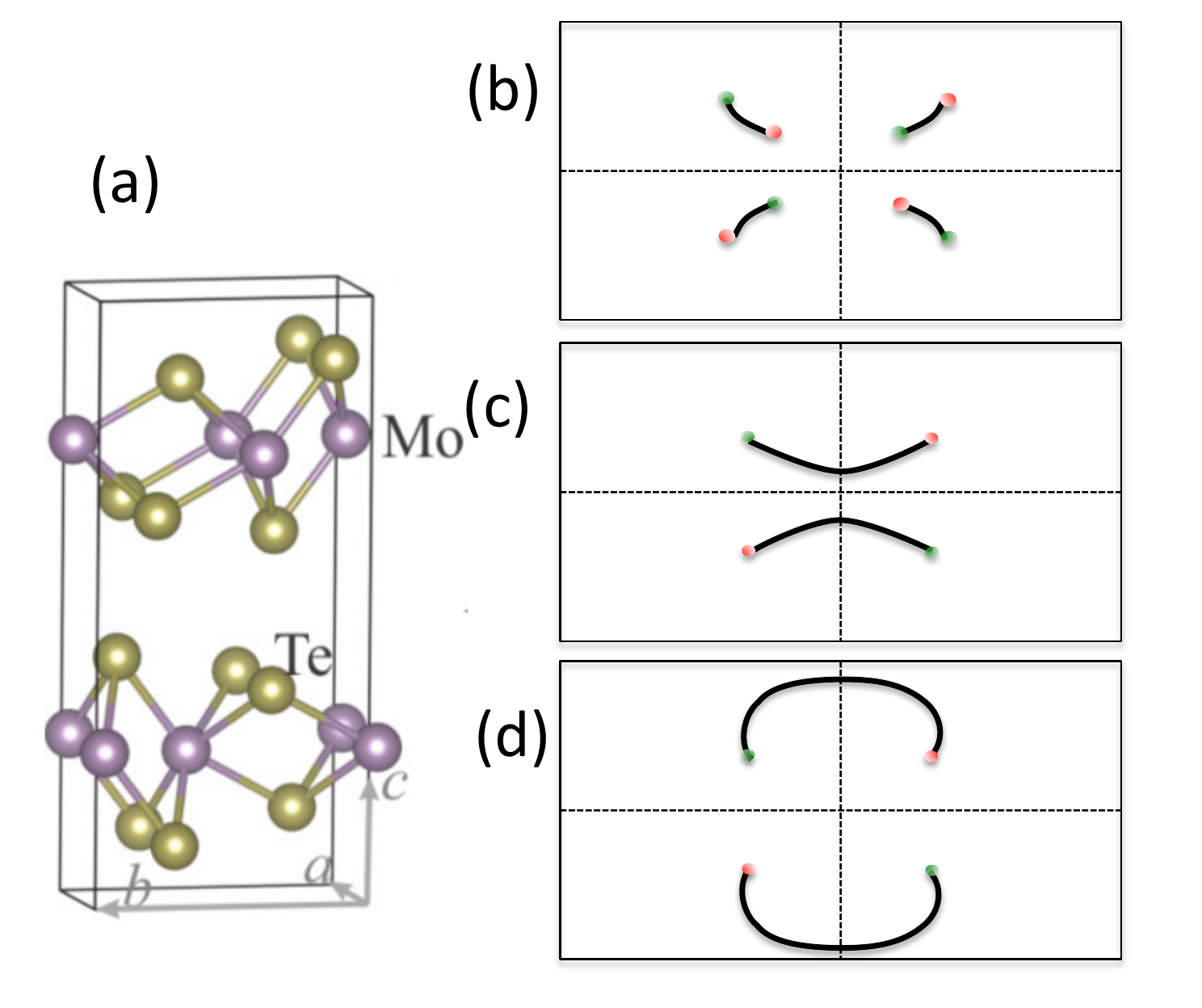}
\label{Figure6}
\caption{Crystal structure and schematics of Fermi arcs for MoTe$_2$. (a) The layered crystal structure ($T_d$ phase) with inversion symmetry breaking. (b)-(d) Three possible scenarios of the Fermi arcs and Weyl points on the surface.
}
\end{figure}

\subsection{Bulk and surface states}
The layered transition-metal dichalcogenide WTe$_2$ was the first theoretically predicted candidate for the type-II WSM~\cite{Soluyanov2015WTe2}. Corresponding Fermi arcs were demonstrated between a pair of Weyl points separated by $\sim$ 0.7\% of the BZ width and located approximately 50 meV above $E_{\mathrm{F}}$, which seems to be challenging for detection using ARPES. Soon, a similar compound, MoTe$_2$~\cite{Sun2015MoTe2}, was proposed to be an optimized version of WTe$_2$, in which the Fermi arcs are six times longer than those of WTe$_2$ and Weyl points are located merely 6 meV above $E_{\mathrm{F}}$. Additionally, W$_x$Mo$_{1-x}$Te$_2$ was also proposed to be a WSM as an alloyed version of the above two compounds~\cite{Chang2016}. WTe$_2$ and MoTe$_2$ crystallize into the same type of orthorhombic lattice without inversion symmetry and show very similar band structures. In the following, we take MoTe$_2$ as an example. 

There are four pairs of Weyl points in the $k_z=0$ plane in the BZ of MoTe$_2$~\cite{Sun2015MoTe2}, as shown in \textbf{Figure 6b}. In the FS, these Weyl points indeed behave as the contact points between the electron and hole pockets. We point out that the band structure is quite sensitive to external strain. Density-functional theory calculations based on slightly shorter lattice parameters revealed that only two pairs of Weyl points exist, while the other two pairs were annihilated by merging at the $\Gamma-X$ line~\cite{Sun2015MoTe2,Wang2015MoTe2}. When these four or two pairs of Weyl points are projected to the surface, Fermi arcs appear to connect each pair of Weyl points with a pair of opposite chirality. Owing to the uncertainty of the sample and experimental conditions, there are three possible scenarios for the connectivity of these Fermi arcs, as demonstrated in \textbf{Figure~6}. The theoretical predictions led to immediate and tremendous experimental activities to verify the type-II WSMs in both MoTe$_2$~\cite{Huang2016,Deng2016,Jiang2016,Liang2016} and WTe$_2$~\cite{Bruno2016,Wang2016WTe2,Wu2016}. However, these works reported different pictures of Fermi arcs including the above three scenarios and even different interpretations, which deserve further investigations before the convergence of conclusions.

\subsection{Magnetoresistance and superconductivity}
The type-I WSM exhibits chiral anomaly for all directions of the chiral anomaly. However, the type-II WSM is argued to show a chiral anomaly only when the magnetic direction is normal to a momentum plane that shows a point-like Fermi surface by intersecting the Weyl point~\cite{Soluyanov2015WTe2}; otherwise, the chiral anomaly effect disappears because the Landau-level spectrum is gapped without chiral zero modes. Longitudinal MR measurements are still called for to address the chiral anomaly in WTe$_2$, MoTe$_2$, and related systems.

It is not surprising that WTe$_2$ and MoTe$_2$ show large transverse MR. Before the discovery of type-II WSMs, WTe$_2$ was already reported for the extremely large MR that increases nearly quadratically with the field~\cite{Ali2014}, for which the topological origin was further proposed more recently~\cite{Muechler2016}. A large transverse MR was also observed for MoTe$_2$~\cite{Keum2015,Qi2016MoTe2}. Interestingly, MoTe$_2$ exhibits superconductivity with a transition temperature of $
T_c = 0.10$ K under ambient conditions~\cite{Qi2016MoTe2}. Under external pressure, both MoTe$_2$ and WTe$_2$ are superconducting with a dome-shaped superconductivity phase diagram, in which the highest transition temperatures are $T_c = 8.2$ K at 11.7 GPa~\cite{Qi2016MoTe2} and $T_c \approx 7$ K at 16.8 GPa~\cite{Kang2015,Pan2015}, respectively.

\section{MULTIFUNCTIONAL HEUSLER MATERIALS}

\subsection{Heusler Topological Insulators}
The Heusler compounds with their great diversity (more than 1500 members) give us the opportunity to search for optimized parameters (e.g., SOC strength, gap size, etc.) across different compounds, which is critical not only for realizing the topological order and investigating the topological phase transitions but also for designing realistic applications~\cite{Graf2011}. In addition, among the wealth of Heusler compounds, many (especially those containing rare-earth elements with strongly correlated $f$-electrons) exhibit rich, interesting ground-state properties~\cite{Yan2014}, such as magnetism~\cite{Canfield1991}, unconventional superconductivity~\cite{Butch2011,Tafti2013,Pan2013,Nakajima2015,Kim2016}, and heavy fermion behaviour~\cite{Fisk1991}. The interplay between these properties and the topological order makes Heusler compounds ideal platforms for the realization of novel topological effects (e.g., exotic particles exhibiting the image monopole effect and axions~\cite{Qi2011RMP}), new topological phases (e.g., topological superconductors~\cite{fu2007c}), and broad applications (see ref. ~\cite{Graf2011} for a review).

Rare-earth Heusler compounds LnPtBi (Ln=Y, La, and Lu) represent a recently proposed model system that can possess topological orders with nontrivial topological surface states and large band inversion~\cite{chadov2010,Xiao2010Heusler,Lin2010}. These compounds have a non-centrosymmetric lattice (space group $F4\bar{3}m$, No. 216). The structure of LnPtBi consists of three interpenetrating $fcc$ lattices; along the [111] direction, the structure can be described as a metallic multilayer formed from successive atomic layers of rare-earth elements, platinum and bismuth.  A band inversion between the $\Gamma_8$ and $\Gamma_6$ bands results in a gapless semimetal with degenerate $\Gamma_8$ bands at the Fermi energy. Because it exhibits an inverted band structure similar to that of HgTe~\cite{bernevig2006d}, LnPtBi is also a TI.

Despite the great interest and intensive research efforts in both theoretical~\cite{chadov2010,Xiao2010Heusler,Lin2010} and experimental~\cite{liu2011,Wang2013Heusler,Shekhar2015Heusler} investigations, the topological nature of Heusler TIs has remained elusive until recently~\cite{Liu2016Heusler,Logan2016}. The Dirac-type surface states have been resolved by performing comprehensive ARPES measurements and $ab~initio$ calculations in the Heusler compounds LuPtBi, YPtBi~\cite{Liu2016Heusler}, and LuPtSb~\cite{Logan2016}. Remarkably, in contrast to many TIs that have topological surface states inside their bulk gap, the topological surface states in LnPtBi show their unusual robustness by lying well below the $E_{\mathbf{F}}$ and strongly overlapping with the bulk valence bands (similar to those in HgTe~\cite{chu2011,bruene2011,Wu2014}). In addition to the topological surface states, numerous metallic surface states were observed to cross the $E_{\mathbf{F}}$ with large Rashba splitting, which not only makes them promising compounds for spintronic application but also provides the possibility to mediate topologically non-trivial superconductivity in the superconducting phase of these compounds.

\subsection{Heusler magnetic Weyl Semimetals}

\begin{figure}[h]
\includegraphics[width=5 in]{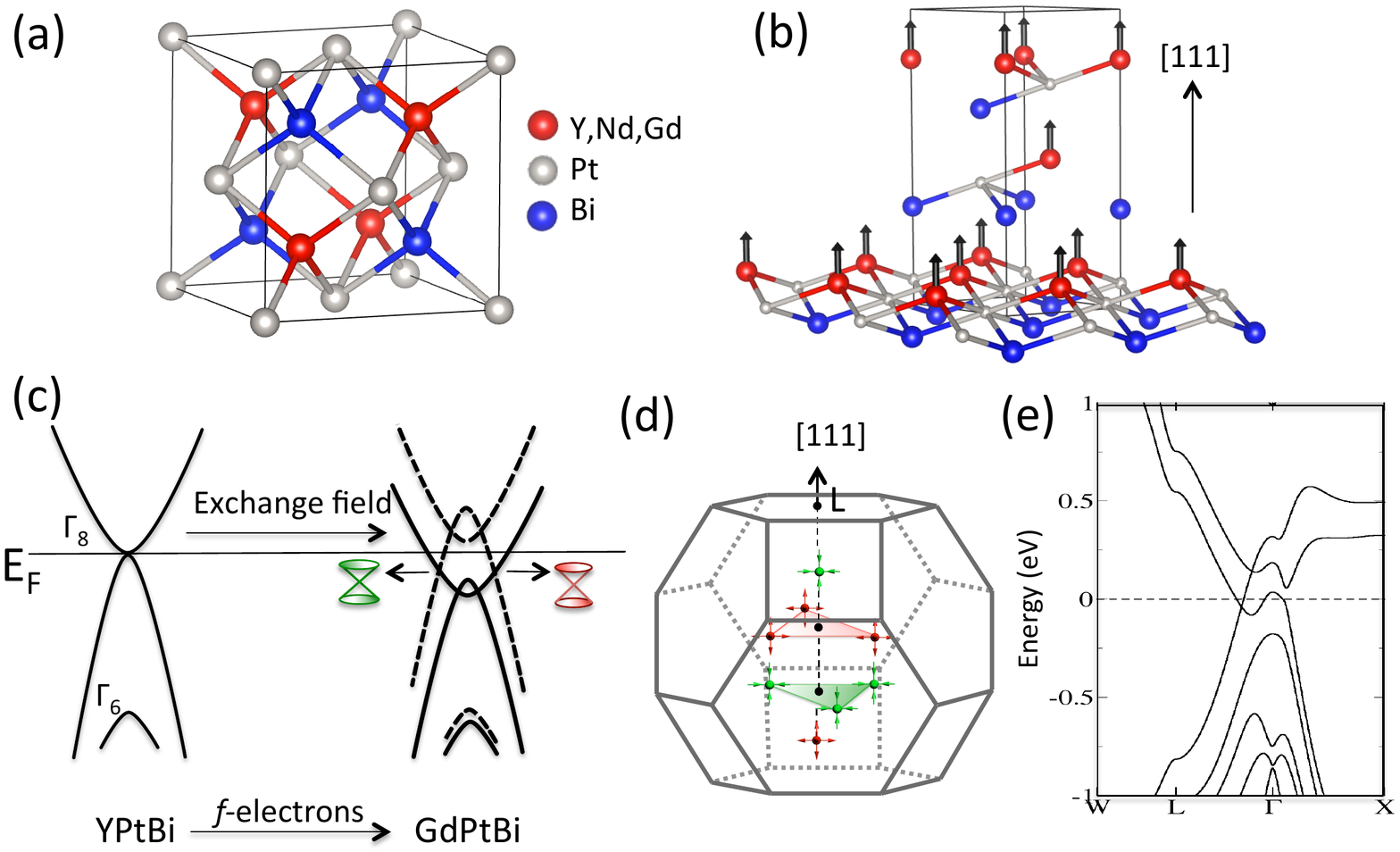}
\label{Figure7}
\caption{Crystal and band structures of Heusler Weyl semimetals. (a) Cubic unit cell of LnPtBi (Ln = Y, Gd or Nd). (b) View of the structure showing Ln-Pt-Bi type layers stacked along the [111] axis. The magnetic moments of the Ln atoms are shown as arrows corresponding to the fully saturated ``ferromagnetic'' state. (c), Schematic comparison of the band structures of YPtBi and GdPtBi. The exchange field from the Gd moments lifts the spin-degeneracy of the $\Gamma_8$ and $\Gamma_6$  bands and induces Weyl points which are slightly below the Fermi energy: the green and red hour-glasses represent Weyl cones with opposite chirality. (D) The distribution of Weyl points in the first Brillouin zone when the Gd magnetic moments are fully saturated along [111] (as shown in (b)). Green and red spheres represent ``--'' and ``+'' chirality, respectively, where the arrows are the Berry curvature vectors. (e) The calculated band structure of GdPtBi (correspond to (c)).
[Adapted from Shekar et al.~\cite{Shekhar2016}]
}
\end{figure}

The combination of band inversion and magnetism in the same Heusler compound provides a way to design the magnetic WSM.
When substituting Ln with most of the lanthanides, the $f$-electrons give rise to magnetism. For example, when Ln = Gd or Nd, GdPtBi and NdPtBi exhibit magnetism arising from their $4f$ electrons, but the $\Gamma_8$ -- $\Gamma_6$ band inversion is preserved. 
 GdPtBi~\cite{Kreyssig2011,Muller2014} as well as NdPtBi~\cite{Muller2015} are antiferromagnetic at low temperatures below their corresponding transition temperatures, $T_N$ =  9.0 K and 2.1 K, respectively. The magnetic structure of these compounds are different: GdPtBi is a type-II antiferromagnet, whereas the magnetic structure of NdPtBi is of type I, indicating that the concrete magnetic ordering below the N$\rm{\acute{e}}$el temperature does not influence the Weyl physics, as we shall see. Additionally, one should be aware of that the size, anisotropy moments, and degeneracy are distinguished for neodymium and gadolinium.  
 
Electronic structure calculations were performed on GdPtBi and YPtBi, as exemplary magnetic and non-magnetic compounds, respectively, to understand the transition from a TI to a magnetic WSM~\cite{Shekhar2016}. A schematic diagram of their energy bands is shown in \textbf{Figure~7}. In GdPtBi, the Gd-$4f$ bands are well localized at energies below the Fermi energy, but we find large exchange-derived spin-splitting of the $\Gamma_8$ and $\Gamma_6$ bands in GdPtBi when there is a net magnetization induced by an external magnetic field, which is absent in YPtBi.  It should be noted that the exchange splitting can be as large as 0.5 eV.
In the presence of an external magnetic field, the magnetization of the Gd moments can be aligned in modest fields, forcing GdPtBi into a ``ferromagnetic'' (FM) state.  
This leads to a pair of Weyl points where the valence and conduction bands touch each other. The orientation of the magnetic moments sensitively affects the Fermi surface and changes the positions and numbers of the Weyl points. For instance, four pairs of Weyl points exist slightly below the Fermi energy when the magnetic moments are along the [111] direction (Fig. 1B), while six pairs of Weyl points appear at different positions in the Brillouin zone when the moments are oriented along the [001] direction~\cite{Shekhar2016}. 

Recently, experimental studies have shown that GdPtBi and NdPtBi become Weyl semimetals when the exchange splitting of the $\Gamma_8$ and $\Gamma_6$ bands is sufficiently large to establish the Weyl points~\cite{Hirschberger2016, Shekhar2016}. This is established for applied fields only of the order of tesla.  It is clear that this is hardly possible from Zeeman splitting in the band structure; rather, the external field results in a significant alignment of the magnetization of the antiferromagnetic structure, resulting in a large exchange field~\cite{Shekhar2016}. Although the magnitude of this exchange field will increase up to the saturation field ($\sim$ 25 T at T = 1.4 K), 
 it is clear that once the exchange field is sufficiently large to induce the Weyl modes, the interesting quantum phenomena of Weyl physics can emerge. 
 Here, GdPtBi and NdPtBi show two strong signatures of the chiral anomaly: a large non-saturated negative quadratic magnetoresistance for fields up to 60 T when a magnetic field was applied parallel to the current direction, and an unusual intrinsic anomalous Hall effect. These signatures, however, are absent in YPtBi, indicating that the $f$ electrons of Gd and Nd must play an essential role in generating the Weyl points. Therefore, it is speculated that all magnetic rare-earth Heusler comounds such as LnPtBi and LnAuSn (Ln = Ce-Sm, Gd-Tm) will show related properties.  
 
 The magnetic field, direction and strength, allows for the tuning of the position and the number of Weyl points in the magnetic Heusler compound.  As another proof of the tunabilty of the band structure, the Seebeck effect is observed to depend strongly on the magnetic field in GdPtBi~\cite{Hirschberger2016}. The Seebeck effect is a voltage generated by a gradient of temperature and intimately determined by the band structure. Therefore, a magnetic field that tailors the band structure sensitively alters the Seebeck effect.

Many other magnetic and non-magnetic Heusler compounds have recently been reported to be WSM candidates, such as the Co-based Heusler materials $X$Co$_2Z$ ($X$  = V, Zr, Nb, Ti, Hf, $Z$ = Si, Ge, Sn)~\cite{Wang2016Heusler} and strained Heulser TI materials~\cite{Ruan2016}.  
Weak TIs~\cite{Yan2012} and nonsymmorphic symmetry protected topological states~\cite{Wang2016} have been reported in the KHgSb honeycomb Heulser materials. Non-Symmorphic Symmetry protected DSMs have also been reported in the antiferromagnetic Heusler material CuMnAs~\cite{Tang2016}.
The chiral antiferromagnetic Heusler compounds Mn$_3X$ ($X$ = Sn, Ge) that exhibit strong anomalous Hall effect at the room temperature~\cite{Kiyohara2015,Nayak2016} 
were also predicted to be antiferromagnetic WSMs~\cite{Yang2016}.
Additionally, several other Heusler-like ternary compounds such as ZrSiS~\cite{Schoop2016} and LaAlGe~\cite{Xu2016Huesler} were also found to be topological semimetals.
Thus, we expect that more WSMs and DSMs will be discovered in the multi-functional and abundant Heusler family and similar compounds.

\section{SUMMARY AND OUTLOOK}

Thus far, several materials have been discovered as Weyl semimetals, in which Weyl points, Fermi arcs, chiral-anomaly-induced negative MR, or the anomalous Hall effect has been demonstrated. However, considerable trivial carrier pockets still coexist with the Weyl pockets in the Fermi surface, which complicates the interpretation of the experimental results. Thus, pure WSM materials are still strongly needed, and only linear Weyl bands appear with the Weyl points located sufficiently close to the Fermi energy. An ideal WSM is expected to exhibit only a single pair of Weyl points at the Fermi energy, in which two Weyl points are well separated in momentum space. 
Current materials are usually in the single-crystal form, while high-quality thin films will be favored for building devices. 

The exotic properties of WSMs show great potential for applications. By exploiting the high mobility and large MR, WSMs can be employed in high-speed electronics and spintronics. We stress that WSMs may exhibit the strong spin Hall effect and related phenomena~\cite{Sun2016} owing to the large Berry curvature and SOC, which is expected to be important for spin Hall effect devices that can efficiently convert charge current to spin current~\cite{Sinova2015}. Finally, one vision for topological insulators and Weyl semimetals is to utilize their robust surface states for surface-related chemical process such as the catalysis~\cite{Muchler2012,Kong2011}. A preliminary theoretical attempt was made to use the surface states of a TI to enhance the activity of traditional catalysts~\cite{Chen2011,Xiao2015} and even reveal topological surface states on traditional catalysts such as gold and platinum~\cite{Yan2015}. The existence of WSM materials have enriched the topological systems that can be considered for the potential design of topological catalysts.

\section*{DISCLOSURE STATEMENT}
The authors are not aware of any affiliations, memberships, funding, or financial holdings that might be perceived as affecting the objectivity of this review. 

\section*{ACKNOWLEDGMENTS}
We are grateful for Yan Sun, Shu-Chun Wu,  Chandra Shekhar, Arnoald Frank, Elena Hassinger, Adolfo G. Grushin and Jens H. Bardarson, Marcus Schmidt, Michael Nicklas, Ajaya K. Nayak, Uli Zeitler, Jochen Wosnitza,  Zhongkai Liu, Yulin Chen and S. S. P. Parkin, for their close collaboration and for their important contributions reviewed in this paper. We acknowledge financial supports from the Max Planck Society and ERC Advanced Grant (291472 ``Idea Heusler'').
%



\end{document}